# Tail Behaviour of the Euro

## JOHN COTTER*
**University College Dublin**

**Address for Correspondence:**
Dr. John Cotter,
Director of Centre for Financial Markets,
Graduate School of Business,
University College Dublin,
Blackrock,
Co. Dublin,
Ireland.
Ph. 353 1 706 8900
Fax. 353 1 283 5482
E-mail. john.cotter@ucd.ie

*The author acknowledges the financial support from University College Dublin President's awards. The author would like to thank Stephen Figlewski, Donal McKillop, Kate Phylaktis and participants of New York's University Salomon Center's conference on the 'Euro: Valuation, Hedging and Capital Market Issues', the European Financial Management Association Annual Meetings and the Statistical and Social Inquiry Society of Ireland for their helpful comments.

## Abstract
This paper empirically analyses risk in the Euro relative to other currencies. Comparisons are made between a sub period encompassing the final transitional stage to full monetary union with a sub period prior to this. Stability in the face of speculative attack is examined using Extreme Value Theory to obtain estimates of tail exchange rate changes. The findings are encouraging. The Euro's common risk measures do not deviate substantially from other currencies. Also, the Euro is stable in the face of speculative pressure. For example, the findings consistently show the Euro being less risky than the Yen, and having similar inherent risk to the Deutsche Mark, the currency that it is essentially replacing.

**Non Technical Summary**

The final transitional stage towards full monetary union for the Euro began on January 1, 1999. This in many ways signalled the true beginning of the currency as thereafter the Euro was allowed for trade in foreign exchange markets and also in the issuance of new public debt. And it is already clear that this new currency will become a serious competitor of the Dollar in international usage. To illustrate, forecasts under a number of economic headings suggest that the Euro will be a closer competitor to the Dollar than having the Deutsche Mark as a vehicle currency (Portes and Rey, 1998). These forecasts have been borne out with a 21% increase in trading volumes for the Euro vis-à-vis the Deutsche Mark from 1998 to 2001 (BIS, 2001). This increase in volume occurred against a backdrop of falling forex transactions. Furthermore, the BIS survey indicates that Euro related trades were 38% of the total market compared to 30% for the Deutsche Mark. Thus, the Euro is a very important currency for the fundamental welfare of the global economy and any concern over its stability is paramount.

It must be remembered that speculative attacks by currency traders have previously unhinged many operating currency systems. Included here was the Mexican Peso disaster that reduced that country's real output by 6 percent in 1995. And also closer to home for participating members of the Euro was the 1992/93 crises of the EU's Exchange Rate Mechanism (ERM), where failed attempts to avert forced devaluations cost all participating members between £100 billion and £150 billion. These two examples show the power of speculators in currency trading and their ability to fundamentally change the value of a currency even against the backdrop of Central

Bank intervention. The issue is whether the Euro exhibited relative large-scale risk compared to other currencies in its final transition stage towards a single currency?

This paper provides a plethora of empirical information pertaining to issues concerning the stability in the newly formed currency. The paper applies Extreme Value Theory which is a methodology that is specifically set up to provide accurate measures of values located at the tail of the distribution of exchange rate changes rather than across the full distribution. We compare the propensity for the Euro (quoted in dollars) to experience extreme exchange rate changes with those of other heavily traded currencies including the Yen, Deutsche Mark, Sterling, Rand and Swiss Franc.

The findings on the whole are very positive towards the Euro. Using a spectrum of quantile and probability levels, we find the Euro as having fluctuations in magnitude between Sterling and the Yen, and very similar to the Deutsche Mark. In essence, this is positive news as the Euro is in effect replacing the Deutsche Mark as the standard European currency vehicle, and there appears to be no extra exchange rate fluctuation problems for the global economy to deal with.

## Tail Behaviour of the Euro

### 1. Introduction

The final transitional stage towards full monetary union for the Euro began on January 1, 1999. This in many ways signalled the true beginning of the currency as thereafter the Euro was allowed for trade in foreign exchange markets and also in the issuance of new public debt. And it is already clear that this new currency will become a serious competitor of the Dollar in international usage. To illustrate, forecasts under a number of economic headings suggest that the Euro will be a closer competitor to the Dollar than having the Deutsche Mark as a vehicle currency (Portes and Rey, 1998). These forecasts have been borne out with a 21% increase in trading volumes for the Euro vis-à-vis the Deutsche Mark from 1998 to 2001 (BIS, 2001). This increase in volume occurred against a backdrop of falling forex transactions. Furthermore, the BIS survey indicates that Euro related trades were 38% of the total market compared to 30% for the Deutsche Mark. Thus, the Euro is a very important currency for the fundamental welfare of the global economy and any concern over its stability is paramount.

To address any concerns requires an examination of the risk inherent in the Euro. The backdrop for this analysis is that participating national currencies of the Euro were previously part of the European Monetary System (EMS) that unfortunately suffered considerably from currency traders' speculative attack on a number of occasions. These attacks reached their zenith during the 1992/1993 crises leading to a radical revamping of the currency system.[1] Turning to the Euro, the inherent instability during the transitional stage no longer exists after full monetary union (Dooley, 1998). This instability results in speculators possibly enjoying opportunities for large trading gains similar to the 1992/1993 currency crises. The issue is whether the Euro exhibited relative large-scale risk compared to other currencies in its final transition stage towards a single currency?

Generally, exchange rate fluctuations are measured unconditionally through an average standard deviation value, or conditionally using a GARCH related specification. However, the objective of this paper is to provide a plethora of information as to whether speculative activity would unhinge participating governments' confidence in the currency in its final transitional stage towards monetary union. To achieve this, the paper concentrates on exchange rate movements located at the tail of the distribution by applying Extreme Value Theory to examine the risk inherent in the newly formed Euro between January 1 1999 and December 31 2001. As the Euro has only recently been traded, we provide extrapolated estimates of the level of risk inherent in the currency.[2] These extrapolated estimates require an examination of, for example, an event that happens every ten years. Extreme Value Theory is optimal from simulation work for a large range of quantile and probability levels as the approach dominates fully parametric and non-parametric approaches in obtaining accurate tail estimates for both in-sample and out-of-sample periods (Danielsson and de Vries, 1997). In particular, using parametric methods where the type of distribution is unknown leads to model misspecification, and a non-parametric approach such as historical simulation involves insufficient data, thereby resulting in poor tail estimates.

Empirically, measures of risk are generally derived by making distributional assumptions, most commonly, normality. However, previous analysis of the unconditional distribution of exchange rates series' has conclusively rejected the assumption of normality (Andersen *et al.*, 1999). Also, a number of stylized facts have been documented as the source of non-gaussian exchange rate series including the unit root property, volatility clustering, and the fat-tailed phenomena (de Vries, 1994). The literature has confronted these issues in a number of ways. First, stationarity is achieved by analysing the first difference of exchange rates rather than the rates themselves (Diebold, 1988). Also, volatility clustering and fat-tails are modelled with a number of alternative distributional hypotheses. These include the mixtures of distributions (Andersen and Bollerslev, 1997), mixed diffusion jump process (Bakaert and Gray, 1996), student-t autoregressive model with dynamic heteroskedasticity (STAR) (Caporale and Pittis, 1996), and many GARCH related processes (Engle and Gau, 1997). However, in the debate for the correct distributional form, there may be a problem in comparing these processes for optimality. Conventional model selection criteria such as comparisons using likelihood ratio tests may incur problems if these distributions are non-nested.

This paper takes an alternative route in detailing the impact caused by large levels of nominal exchange rate changes. It applies methods evolving from Extreme Value Theory where it is unnecessary to have full knowledge of the unconditional distributional form. This theoretical framework does not try to model the full distribution of the variable of interest, but rather, it deals specifically with the values located at the tail of the distribution. As this paper is going to make judgements as to the comparability of large exchange rate fluctuations for the Euro with other currencies, it is vital that we concentrate on tail estimates. Also, forecasting whether a system, for example the Euro, may incur too much risk and thus have stability problems requires an analysis of the extreme exchange rate fluctuations.

The rest of the paper proceeds as follows. In the next section, Extreme Value Theory is presented detailing the method used in tail estimation and the related procured risk measures. The advantages of applying Extreme Value methods in the context of measuring the impact of the exchange rate tail estimates are also outlined. In section 3 a preliminary description of the Euro and comparative currencies is given. Results from unit root tests as well as unconditional standard deviation and conditional volatility measures are presented. Section 4 presents the main empirical findings. First the semi-parametric tail estimates and their implications are outlined. Then, the findings for the two Extreme Value measures, the quantile and excess probability estimators are discussed. Here the issue of the stability of the Euro is analysed. Finally a summary of the paper and some conclusions are given in section 5.

## 2. Theory and Methods

### 2.1. Extreme Value Theory

Extreme Value Theory is used to calculate the impact of nominal exchange rate changes located at both the upper and lower tails of the distribution of all currency changes. In its purest form, a random variable R is assumed to be independent and identically distributed (iid) and belonging to the true unknown cumulative probability density function F. Less restrictive assumptions are equally applicable through the maximum domain of attraction (MDA) which allows for approximation to certain distributional characteristics rather than belonging to a specific distribution (Leadbetter *et al.*, 1983). The case of currency first differences would only be able to rely on the assumption of stationarity rather than iid and these requirements are considered analogous for the purposes of this paper. Dealing with a sequence of exchange rate changes arranged in ascending order and expressed in terms of the maxima ($M_n$) of n random variables where

$$M_n = \max \{R_1, R_2,..., R_n\} \qquad (1)$$

The corresponding density function of $M_n$ is got from the cumulative probability relationship and this represents the probability of exceeding a certain return on the upside distribution of exchange rate changes for n first differences:

$$P_{positive} = P\{M_n > r_{positive}\} = P\{R_1 > r_{positive}, ..., R_n > r_{positive}\} = 1 - F_{max}^n(r_{positive}) \qquad (2)$$

$r_{positive}$ represents the exchange rate change located on the upside of a distribution currency changes.

However, the random variables of interest to us are located at both upper and lower tails of the distribution $F^n(r)$ and Extreme Value Theory is usually notated for upper order statistics. The theory is equally applicable for lower order statistics, as in the specific case of downside risk measures. Lower tail rate movements are relevant in this analysis of values located on the downside distribution of all exchange rate changes for n first differences. For convenience purposes, the expressions used in this paper are in terms of upper order estimators. The theoretical framework for examining sample minima tail statistics can easily be converted by applying the identity $\text{Min}\{R_1, R_2,..., R_n\} = -\text{Max}\{-R_1, -R_2,..., -R_n\}$. The corresponding probability expression for exceeding a certain return on the downside distribution of exchange rate changes:

$$P_{negative} = P\{M_n < r_{negative}\} = P\{R_1 < r_{negative}, ..., R_n < r_{negative}\} = F_{min}^n(r_{negative}) \qquad (3)$$

$r_{negative}$ represents the exchange rate change located on the downside of a distribution of exchange rate changes.

The rest of the Extreme Value Theory theory framework is presented for the positive first differences  outlined in (2) but it is equally applicable for negative first differences.

### 2.1.1. Asymptotic Behaviour of Distribution

While the exact distribution of the n random variables of (1) may not be known, we rely on the Fisher-Tippett theorem to examine asymptotic behaviour of the distribution.  From this theorem, there are three types of limit laws and these are what make up the Extreme Value distributions.  Formally, taking the sequence of n random variables from the non-degenerate distribution function H, and assuming that the distribution of F converges at the limit to the distribution, these belong to the following types of distributions:

Type I (Gumbell): $\Lambda(r)$        $= \exp[-e^{-r}]$            $-\infty < r < \infty$

Type II (Fréchet): $\Phi_\alpha(r)$      $= 0$                 $r \leq 0$

                                  $= \exp[-r^{(-\alpha)}] = \exp[-r^{(-1/\gamma)}]$      $r > 0$

Type III (Weibull): $\psi_\alpha(r)$    $= \exp[-(-r)^{(\alpha)}] = \exp[-(-r)^{(1/\gamma)}]$      $r \leq 0$

                                  $= 1$                            $r > 0$          (4)

and for $\alpha > 0$.

Where $\alpha$ measures the tail shape of the associated parent distribution.

The above distributions belong to the set of max-stable distributions that are analogous to the Extreme Value distributions.  Max-stable distributions are said to belong to a class of all possible non-degenerate limit laws for the maxima of n random variables.  Thus, the max-stable property permits the classification of all limit laws under the three headings in (4). These limit laws indicate weak convergence, that is convergence in distribution rather than in probability. Otherwise convergence would not take place for all iid random variables. As such, the convergence of the maxima of n random variables is analogous to convergence of the sums given by the central limit theorem, as the asymptotic convergence in probability is not equal to 1. Further details of the limit laws and asymptotic convergence are available in the appendix.

All moments are well-defined for the Gumbell and Weibull distributions in contrast to the Fréchet distribution. These distributions can be divided into three separate groups depending on the value of the shape parameter $\alpha$ in (4). The classification of a Weibull distribution ($\alpha < 0$) includes the uniform example where the tail is bounded by having a finite right end point and is a short tailed distribution. The more commonly assumed class of distributions used for exchange rate changes includes the set of thin tailed densities, and most notably amongst these, the normal or lognormal distributions. This second classification of densities includes the generated normal and exponential distributions and these belong to the Gumbell distribution, having a characteristic of tails decaying exponentially. Of primary concern to the analysis of fat-tailed distributions is the Fréchet classification, and examples of this type generated here are the Cauchy, student-t, ordinary fréchet, and the pareto distributions. This important classification of distributions for extreme exchange rate movements has tail values that decay by a power function. The power decline results in the Fréchet distribution declining slower than the Gumbell distribution.

A vast literature on nominal exchange rate changes (Dewachter, 1995; Koedijk and Kool, 1994; Koedijk *et al.*, 1992; and Diebold, 1988) has recognised the existence of fat-tailed characteristics. The necessary and sufficient condition using Gnednenko's theorem for a distribution to asymptotically converge on the (fat-tailed) Fréchet type of Extreme Value distributions is:

$$\text{Type II (Fréchet):} \qquad \lim_{t \to \infty} \frac{1 - F(tr)}{1 - F(t)} \quad = \quad r^{-1/\alpha} \quad = \quad r^{-\gamma} \qquad (5)$$

This condition allows for unbounded moments and represents a tail having a regular variation at infinity property (Feller, 1971). Practically a distribution of random numbers with the property of regular variation implies that the unconditional moments of r larger than $\gamma$ are unbounded.

### 2.1.2. Maximum Domain of Attraction

Once one assumes that exchange rate changes have fat-tailed characteristics, we can infer that they are in the maximum domain of attraction (MDA) of a Fréchet type distribution. This implies that the exchange rate changes do not exactly fit a particular set of distributional assumptions, but rather, similar to the central limit theorem, we assume that sequences of stationary exchange rate changes have extreme values that are approximated by the Fréchet type Extreme Value distribution. Whereas, the estimation of exact fits to a particular distribution encourages parametric assumptions and estimation techniques, measuring approximations to distributions utilises semi-parametric frameworks. Also, from an analysis of different extremal statistics, Semi-parametric measures offer an advantage over their parametric counterparts in that under non-gaussian conditions one obtains better bias and mean squared error properties (Danielsson and de Vries, 1997). Formally we can denote the characteristic of belonging to the maximum domain of attraction (MDA) as

$$R_1, R_2, ..., R_n \text{ are stationary from } F \in \text{MDA } (H_\gamma) \tag{6}$$

And in the specific case of a Fréchet distribution approximation, (6) reduces to

$$F(r) = r^{-\alpha}L(r), \qquad r > 0 \tag{7}$$

Where $\alpha$ has parametric assumptions, whereas $L(r)$ is some slowly decaying function that is underpinned by semi-parametric assumptions. While there is a general agreement on the existence of fat-tails for exchange rate changes, its exact form for all financial first differences is unknown. For this reason, it is appropriate to deal with approximation of the Fréchet distribution in the sense of being in the maximum domain of attraction.

### 2.2. Quantile and Probability Estimators

Due to the semi-parametric specification of being in the maximum domain of attraction of the fat-tailed Fréchet distribution, it is appropriate to apply semi-parametric measures of our tail estimates. In the literature a number of moment tail estimators based on order statistics have been suggested of which the moments based Hill (1975) estimator dominates in terms of bias and efficiency (Kearns and Pagan, 1997). Dealing specifically with the highest extreme exchange rate changes, the estimator is applied to the sequence placed in ascending order:

$$\gamma_h = 1/\alpha = [1/m] \sum [\log \{r_{(n+1-i)}\} - \log \{r_{(n-m)}\}] \qquad \text{for } i = 1....m \tag{8}$$

The Hill tail estimator is asymptotically normal, $(\gamma - E\{\gamma\})/(m)^{1/2} \approx (0, \gamma^2)$ (Hall, 1982).

As this study is examining the probability of a sequence of exchange rate changes exceeding a particular exchange rate change relying on expressions (2) and (3), an empirical issue arises in determining the number of exchange rate changes entailed in the tail of a distribution. Amongst the methods for finding the optimal threshold of where the tail of a distribution begins, we adopt the approach initially developed by Hall (1982) and updated by Phillips $et\ al.$ (1996). The optimal threshold value, $M_n$, which minimises the mean square error of the tail estimate, $\gamma$, is $m = M_n = \{\lambda n^{2/3}\}$ where $\lambda$ is estimated adaptively by $\lambda = \left| \gamma_1/2^{1/2}(n/m_2(\gamma_1 - \gamma_2) \right|^{2/3}$. The preliminary estimates of $\gamma_1$ and $\gamma_2$ are obtained using (8) with data truncations $m_1 = n^{\sigma}$ and $m_2 = n^{\nu}$. Phillips $et\ al.$ (1996) suggest using $\sigma = 0.6$ and $\nu = 0.9$ as the values of m and $\gamma_1$ are insensitive to the choice of $\sigma$ and $\nu$ for small intervals around these estimates.

On obtaining the tail estimate we can make a number of conclusions regarding the $\gamma$ coefficient obtained for the series of nominal exchange rates analysed. Koedijk $et\ al$. (1992) find that floating rates should have a greater coefficient than their fixed counterpart. Floating rates involve exchange rate setting through market makers alone, whereas a fixed system will allow for a combination of market makers and government intervention (in the case of very large fluctuations). The fluctuations will tend to be smoother for the floating rates due to the lack of large changes caused by large-scale government intervention. Thus, while floating rates involve a greater degree of risk on average, the impact of realignments will induce the fixed rates to have relatively more extreme fluctuations and these result in fatter tails. So, the probability mass in a tail and the tail estimate are negatively related, that is the fatter the tails, the lower the $\gamma$ coefficient. Also, we can infer whether the series under analysis displays a finite variance according to its Hill estimate. For example, an infinite variance is in line with γ less than 2, whereas a value greater than this implies the existence of the second moment. Thus, we can differentiate between degrees of fat-tailed distributions ranging from the stable process to a student-t density with our methods.

Finally, we can determine if the extreme behaviour of nominal exchange rate changes remains constant for its upper and lower limits using (9). Take the downward pressure on the Euro during 1999 as an example, this may involve extreme values. Thus policy makers may be more concerned with downside Euro rate movements than their upside counterparts, and they may create a fund for intervention purposes that is specifically for use against this type of speculation. In order to investigate whether the downside tail measures are similar or not to the upside measures, the tail index estimator is used to determine each tail individually, and also to measure a common value encompassing the extreme rate movements of both tails. Equation (8) is set up to deal with the positive exchange rate changes, whereas a simple rearrangement of the sequence of exchange rate changes focuses on large negative exchange rate changes, and using absolute values encompasses all currency fluctuations. The relative stability of the tail measures points out any differences in the extent of extremities in the tails. Tail stability is tested using a statistic suggested by Loretan and Phillips (1994):

$$V(\gamma^+ - \gamma^-) = [\gamma^+ - \gamma^-]^2 / [\gamma^{+2}/m^+ + \gamma^{-2}/m^-]^{1/2} \qquad (9)$$

for $\gamma^+$ ($\gamma^-$) is the estimate of the right (left) tail.

This test statistic is also applied to tail values for preEuro and postEuro periods to determine the extent of any tail behaviour changes (if any) that occurred from switching to the final transition stage for the single currency.

Using (8) to determine the tail index, various quantile estimates can be generated using the following:

$$r_p \qquad = \qquad r_t(m/np)^\gamma \qquad\qquad\qquad (10)$$

The quantiles measured by (10) can be used to generate different exchange rate changes for various confidence levels. A related measure requiring the calculation of the tail index is the Excess Probability estimator given in (11).

$$P_r \qquad = \qquad (r_t/r_m)^{1/\gamma}m/n \qquad\qquad\qquad (11)$$

The Excess Probability estimator shows the probability of having a nominal exchange rate change exceed a certain quantile threshold point, $r_t$.

### 3. Preliminary Findings

Officially the Euro replaced the participating Exchange Rate Mechanism's member's own currency on January 1 1999. As well as analysing the Euro independently, five separate currencies quoted in Dollars are chosen for comparison purposes. These are the Deutsche Mark, Sterling, Yen, Rand, and Swiss Franc. The time period of analysis is January 1 1990 to December 31 2001 incorporating two sub periods, preEuro (1990-1998) and postEuro (1999-2001).[3] The latter represents the final transition stage towards full monetary union for the Euro's participants and is the main focus of examination. A time series plot of the Euro's log exchange rate is presented in figure 1 where we see the strong depreciation in the initial stages of the new currency relative to the Dollar. Analysis is based on the log rate to avoid technical problems, namely Jensen's inequality.[4] These log variables are also first differenced to examine exchange rate changes to allow for unit-free comparisons, and more importantly, because of the unit root property. Augmented Dickey Fuller (ADF) and Phillips Perron (PP) unit root tests are presented for the spot rates in table 1. The findings are consistent with past studies modelling exchange rates, namely that log exchange rates series' do not accept the hypothesis of stationarity (see Diebold, 1988). Following convention, the exchange rate series' are first differenced to avoid spurious conclusions. In table 1 we see that the use of exchange rate first differences in all cases result in stationary series'.

INSERT FIGURE 1 HERE

INSERT TABLE 1 HERE

Figure 2 gives an early indication of the magnitude of fluctuations in the Euro series. Most, but not all, daily first differences are within a one percent band. Table 2 gives more concrete evidence of the characteristics of the Euro where comparable preliminary statistics for all the exchange rate series are presented. The Euro, similar to other currency first differences, exhibits excess kurtosis and non-normality. Excess kurtosis is often used as the signal for fat-tails as the property implies that there is a bunching of first differences around the tails of a distribution. Like the Rand, the Euro is significantly positively skewed since 1999. Similarly, all currencies except the Deutsche Mark indicate significant positive asymmetries before 1999. The preliminary indication of volatility for the Euro is that the measure is quite substantial using unconditional standard deviation, with an average daily volatility in excess of 0.5 percent. However all values for dispersion including the range and interquartile range measures are similar for the Deutsche Mark pre 1999.[5] In contrast, these measures are generally greater in magnitude for the Yen and the Rand.

INSERT FIGURE 2 HERE

INSERT TABLE 2 HERE

Prior to detailing the Extreme Value estimates, the conditional volatility for the Euro is measured using a GARCH (1, 1) model (Bollerslev, 1986). This model was chosen from a host of successful candidates for modelling exchange rate fluctuations including GARCH-M (McCurdy and Morgan, 1988), SWARCH (Fong, 1998), Absolute Value GARCH (Neely, 1999), and GARCH-MA (Engle and Gau, 1996) specifications. Importantly, given the basis of this paper focuses on the tail behaviour of exchange rate first differences, a GARCH (1, 1) model adequately describes volatility clustering which leads to a fat-tailed property in currency first differences (Hsieh, 1989). In fact, this property has long been identified for financial time series (for currency series see Mussa, 1979). For example, the 'Noah Effect' observes that "large changes tend to be followed by large changes – of either sign – and small changes tend to be followed by small changes" (Mandelbrot, 1963). The resulting impact of this phenomenon is that high levels (and low levels) of volatility cluster together.

Turning to the dynamics, accurate measures of time varying volatility as proxied by its variance, $h^2_t$, are estimated with a GARCH (1, 1):[6]

$$h^2_t = \alpha_0 + \sum_{i=1}^{q} \alpha_i \varepsilon^2_{t-i} + \sum_{j=1}^{p} \beta_j h^2_{t-j} \qquad (12)$$

Figure 3 represents a plot of the conditional standard deviation for the Euro with the related models coefficients. The model appears to be well specified with both significant autoregressive and moving average variables. Furthermore, ARCH effects are removed as indicated by the Ljung-Box test on the standardised residuals (6.319), and on the squared standardised residuals (7.345) where both test statistics indicate insignificant second moment dependence. The time-varying characteristic of the conditional volatility of the Euro is evident. These fluctuations resulted in a large peak around January 2001. In the next section, we document accurate measures of unconditional large-scale risk.

INSERT FIGURE 3 HERE

## 4. Extreme Value Estimates

### 4. 1 Tail Estimates

The results presented in this section focus directly on the large-scale fluctuations inherent in the nominal exchange rate series. From this analysis, conclusions are made regarding the risk of the Euro in comparison to that of other currencies. In order to obtain quantile and excess probability estimates, we first obtain tail estimates for the exchange rate series' and these are presented in table 3. Estimates for the left, right and a measure encompassing currency movements for both tails are presented. A number of interesting observations can be made regarding table 3. First for each sub period, the Hill estimates remain stable across all tails except the Yen PRE1999 using Loretan and Phillips (1994) stability test, $\gamma^+ - \gamma^-$. The extent of any divergence is detailed for the five largest extreme exchange rate changes in table 4. In general the values are reasonably similar except the Yen PRE1999 and the Rand POST1999. While it initially appears that negative values are greater in magnitude than positive extreme values, this pattern is not statistically significant. Thus further discussion concentrates on the combined tail measure, $\gamma^*$. Intuitively, this is the variable that policy makers would be interested in, as they are concerned with the magnitude in all currency fluctuations *per se*, rather than the specific case of upside or downside risk.

INSERT TABLE 3 HERE

Second, the Euro exhibits similar tail values to the other series with the exception of the Rand. Statistically, this difference is significant (t-statistic 2.71), although similarly the Deutsche Mark is also statistically diverging in the preEuro period (t-statistic 4.09). Thus, the Euro represents more of a floating currency than the Rand as higher $\gamma^*$ implies a smoother processing of extreme values.[7] For the Euro, the magnitude of very extreme values is reduced with market equilibrium decided only by market makers without any realignment by the authorities. Comparing the results in table 2 and 3, we see a combination of currencies having higher average volatility as measured by standard deviation and other measures of dispersion, and having fatter tails where the magnitude of extreme values is greater. Taking the Euro and the Rand as an example, we see in table 2 that the Euro exhibits relatively lower levels of average volatility during the period of analysis. In addition, the Euro has a higher tail estimate indicating smaller discrete rate jumps over time as supported by the most extreme exchange rate changes in table 4.

INSERT TABLE 4 HERE

Third, we can classify the existence of moments for a distribution based on the calculated Hill values. Our empirical estimates change from the largest value of 3.82 to the smallest value of 2.16 occurring in the preEuro period, whilst it has been noted that a finite variance and kurtosis have values of 2 and 4 respectively (Loretan and Phillips, 1994). With our estimates, we find that the hypothesis of a second moment is acceptable, whereas a finite fourth moment is not. Formally, this is examined using a difference in means test statistic. All exchange rate series' with the exception of the Rand exhibit a finite variance as rejection of the hypothesis requires that the test statistic be in excess of -1.64. In contrast, there is no support for the existence of the finite kurtosis coefficient as the null hypothesis, $H_o$: $\gamma \geq 4$, is only accepted if the test statistic is greater than 1.64. In fact it is formally rejected for the Yen and the Rand for both subperiods with test statistics in excess of $-1.64$.

Fourth the tail values are used to discern between association with different distributions. For example, Hill estimates with values between 1 and 2 have stable paretian characteristics of which the cauchy and normal distributions have values of 1 and 2 respectfully. In contrast, fat-tailed distributions such as the student-t(3), have values in excess of 2 (Ghose and Kroner, 1995). Formally this is tested using a difference in means statistic with the null, $H_o$: $\gamma \leq 2$, representing membership of the stable paretian family of distributions, and the alternative, $H_a$: $\gamma > 2$, offering support for fat-tailed characteristics for all exchange rate series' analysed except the Rand for the two subperiods and the Yen POST1999 (for example, 2.08 for the Euro). Also, there is no conclusive evidence of support for the Rand being part of the stable paretian family of distributions.

Of primary focus in this study is to determine whether the introduction of the Euro has involved a significant change in tail behaviour. This issue is examined using the stability test proposed by Loretan and Phillips (1994) and results are presented in table 4. Hill tail estimates, $\gamma^*$, are compared for PRE1999 and POST1999 with positive (negative) values representing an increase (decrease) in tail estimates over time. Three important conclusions can be extracted by the stability tests. First, there are (in some cases) significant changes in tail estimates across currencies over time. In total there are eight significant changes involving five negative (reductions in tail estimates) and three positive ones. However, all of these changes except one involve the Rand as a currency pair and as noted in table 3, this currency has very low tail estimates vis-à-vis the other currencies over the subperiods.

INSERT TABLE 5 HERE

Second, and more importantly, there is no significant divergence for a currency over time as indicated by the diagonal of the matrix of values in table 4. Whilst there are small changes in the tail estimates, this lack of statistical significance implies that tail behaviour for each currency remains reasonably stable over time. Finally, and the key focus of this analysis is the rejection of the hypothesis that the introduction of the Euro involved a systematic change in tail behaviour compared to the Deutsche Mark. Although we do see a reduction in the tail estimates post 1999 this is not significant (test statistic –0.20). An economic interpretation of this finding is that the introduction of the Euro did not result in any fundamental view change from economic agents compared to the Deutsche Mark, the currency that it is essentially replacing. Furthermore, the stability across time implies that for the new currency in general extreme exchange rate changes should not occur with more frequency.

*4.2 Quantile and Excess Probability Estimates*

Using the Hill tail values, quantile estimates are presented in table 6. Taking the Euro $r_p$ (95) estimate as an example, the quantile implies that ninety five percent of all nominal currency fluctuations are approximately within a one percent (1.15) band. As expected, the fluctuations increase as you move to a more extreme quantile. For example, by moving along the probability grid from ninety five to ninety nine percent increases the exchange rate first differences in the Euro to 1.78%. The last three columns represent quantile values for a probability level that is dependent on the sample size, n, of the currency series analysed. These estimates are attractive for extrapolation to out-of-sample estimates for the Euro. For example, $r_p$ (1/4n) for the Euro deals with 3104 days (776 * 4), and the related probability level calculated by 1/4n. Thus, approximately every twelve years (3104 days), an exchange rate change of 4.49% should occur. This represents an estimate of the risk that the Euro is capable of over a twelve-year period.

INSERT TABLE 6 HERE

In table 6, we see that exchange rate changes for the Euro are very similar to the Deutsche Mark and Sterling, although they tend to dominate the latter currency. In contrast, the more volatile Yen and Rand exhibit greater exchange rate changes at different quantiles. Comparing the extrapolated estimates, we see that for a twelve-year time frame, the Rand is capable of having fluctuations of over nine percent vis-à-vis the Euro with less than five percent. Also, there is no quantile level where the magnitude of the exchange rate change for the Euro exceeds the Rand.

Similar conclusions can be made from analysing table 7. Here, the second Extreme Value measure, the excess probability estimate, $P_r$, deals with the issue of having exchange rate fluctuations in excess of a certain percentage. The findings show that the probability of having an exchange rate change in excess of 5% to be very low, for example, 0.02 in the case of the Euro. As we would expect *a priori*, the probability of exceeding a certain value increases as you move to smaller fluctuation levels.

INSERT TABLE 7 HERE

The information in table 7 can be used to make decisions on the basis of statistical inference. Whilst, much of the creation of the Euro was down to political decision making, we can now make some comments about the Euro on its final stage prior to becoming a single currency that are based on the statistical evidence of extreme exchange rate changes. First, the probability of having very large exchange rate changes is more likely for the Yen and the Rand than for the Euro. Second, for comparatively smaller currency fluctuations, the probability of their occurrence is again larger for the Yen and the Rand than for the Euro.

Most importantly, Dooley (1998) suggests that whilst successful speculative attack causing the abandonment of the Euro is very unlikely after its full implementation, the probability of such an outcome during the transitional stage is higher.[8] This conclusion is due to very high costs that would have to be incurred by participating governments on abandonment after full monetary union enforcing these agents to decline carrying out this decision. Also for this regime, there are no gains to the government as they have consistently repaid debts outstanding after abandoning exchange rate mechanisms. In contrast, abandonment of the transitional movement towards the Euro would cost the government in terms of the speculators' profits from, for example, devaluation, but would also reduce the magnitude of domestic currency denominated liabilities outstanding. Therefore, there is some benefit to this action in the transitional stage.

This leads to two related questions assuming that policy makers make decisions on the basis of statistical inference from risk levels alone. First, is there any likelihood that the Euro would suffer inherent instability due to excess risk? Second, should we be worried about the levels of fluctuations in the Euro? Taking the Yen as a comparable benchmark, the answer to this first question is negative. As mentioned, the Yen yields a greater probability of those fluctuations occurring that causes concern about the stability of a currency than the Euro, namely the movements in excess of 5 or 3 percent. In response to the second question, we also report low probabilities of having relatively smaller fluctuations in excess of 1 percent movements. Thus, taking these two issues together, it is reasonably safe to assume that the Euro has shown all the characteristics of a floating currency and that while daily volatility is material in size, there are no fears of any fundamental problems with the currency.

## 5. Summary and Conclusion

This paper uses Extreme Value Theory to determine the impact of tail exchange rate first differences for a number of currencies including the Euro. This new currency is seen as a major player in its role as a prime international vehicle currency. Given previous currency crises and their impact on the global economy considerable attention has been placed on the prospects of success for the Euro during its transition to full monetary union. As is known, intensive speculative trading results in major fluctuations and undermines confidence in a currency. So stage 3 of monetary union between 1999 and July 2002 is a crucial time for these economies to analyse the levels of risk that the Euro is capable of, as thereafter the probability of speculative attacks is reduced.

Previous analysis of exchange rate series has found a non-stationary property, volatility clustering and a fat-tailed distribution. The Euro is no different, as it requires first differencing to attain stationarity, the fat-tail characteristic is indicated with the measure of kurtosis and tail estimates, and volatility clustering is implied by our GARCH modelling. Given our focus on the tail first differences, Extreme Value Theory and the fat-tailed limiting distribution, namely the Fréchet distribution, are utilised in the calculation of extreme return measures. Tail estimates are given using the semi-parametric Hill estimate that accounts for the relationship that exists between exchange rate first differences and the Fréchet distribution through the maximum domain of attraction. The Euro's tail behaviour is found to be no different form the Deutsche Mark indicating high levels of stability in the face of speculative attack.

Statistical inference is made on two important questions using our estimated extreme return measures. First, is there any concrete likelihood that the extent of the risk in the Euro would lead to inherent instability thereby attracting intensive speculative attacks? Second, are the fluctuations inherent in the Euro actually a cause for concern? We answer both of these questions in the negative as we see that exchange rate changes in the Euro are of a similar magnitude to other floating currencies, and in fact smaller than exhibited by the Yen. Using a spectrum of quantile and probability levels, we find the Euro as having fluctuations in magnitude between Sterling and the Yen, and very similar to the Deutsche Mark. In essence, this is positive news as the Euro is in effect replacing the Deutsche Mark as the standard European currency vehicle, and they appear to be no extra exchange rate fluctuation problems for the global economy to deal with.

Figure 1: Plot of Daily Euro Log Exchange Rates

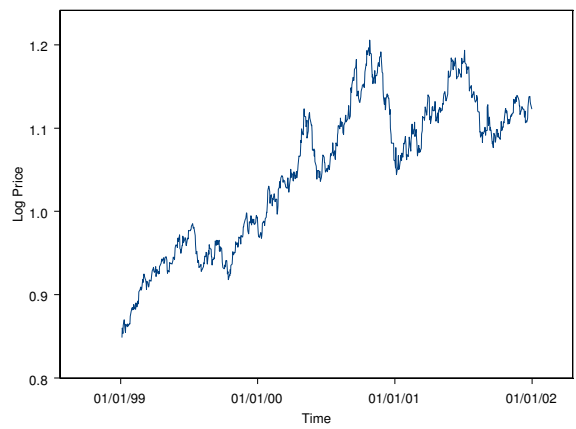

Notes: The figure details the Euro log exchange rates in terms of one unit of a US Dollar.

Table 1: Stationarity Tests for Exchange Rate Series

| Currency | Dickey Fuller (Levels) | Phillips Perron (Levels) | Dickey Fuller (Returns) | Phillips Perron (Returns) |
|---|---|---|---|---|
| **PRE1999** | | | | |
| **Deutsche Mark** | -2.37 | -10.37 | -17.67* | -2277.49* |
| **Yen** | -1.61 | -4.16 | -12.59* | -2369.12* |
| **Sterling** | -2.24 | -10.78 | -17.19* | -2404.84* |
| **Rand** | -2.57 | -11.74 | -9.57* | -2359.39* |
| **Swiss Franc** | -2.35 | -9.83 | -28.78* | -2311.99* |
| | | | | |
| **POST1999** | | | | |
| **Euro** | -2.32 | -11.47 | -16.58* | -831.48* |
| **Yen** | -1.15 | -3.79 | -17.11* | -833.12* |
| **Sterling** | -2.58 | -14.37 | -17.02* | -814.63* |
| **Rand** | 0.95 | 3.69 | -6.90* | -859.67* |
| **Swiss Franc** | -2.26 | -11.06* | -17.06* | -837.75* |

Notes: All log exchange rates are in terms of one unit of a US Dollar. The Euro replaces the Deutsche Mark for the POST1999 sample period. Both tests are included for technical reasons. The Phillips Perron test is the semi-parametric counterpart to the Augmented Dickey Fuller test, and exchange rate series' indicate semi-parametric characteristics. For each series, the optimum number of augmenting lags is chosen on the basis of the Akaike Information Criteria. The symbol * represents values in excess of the critical values found in Phillips and Ouliaris (1990).

Figure 2: Plot of Daily Euro Returns

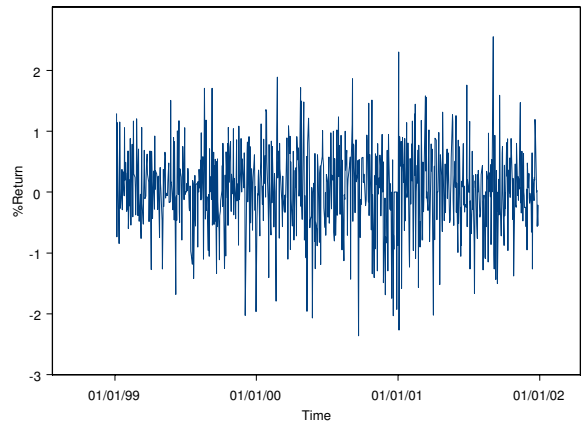

Notes: The figure details daily Euro percentage first differences in terms of one unit of a US Dollar.

Table 2: Summary Statistics for Exchange Rate Returns Series

| Currency | Mean | Standard Deviation | Range | Interquartile Range | Skew. | Kurt. | Normality |
|---|---|---|---|---|---|---|---|
| **PRE1999** | | | | | | | |
| **Deutsche Mark** | -0.001 | 0.680 | 6.962 | 0.703 | 0.020 | 2.392* | 0.056* |
| **Yen** | -0.010 | 0.749 | 11.825 | 0.721 | 1.110* | 9.496* | 0.084* |
| **Sterling** | -0.001 | 0.647 | 8.507 | 0.620 | 0.170* | 4.577* | 0.078* |
| **Rand** | 0.035 | 0.512 | 8.034 | 0.332 | 0.512* | 14.595* | 0.139* |
| **Swiss Franc** | -0.004 | 0.761 | 7.487 | 0.807 | 0.121* | 2.182* | 0.060* |
| | | | | | | | |
| **POST1999** | | | | | | | |
| **Euro** | 0.036 | 0.691 | 4.912 | 0.797 | 0.243* | 0.740* | 0.048* |
| **Yen** | 0.021 | 0.704 | 6.804 | 0.792 | 0.016 | 3.089* | 0.054* |
| **Sterling** | 0.016 | 0.503 | 3.491 | 0.567 | 0.114 | 0.715* | 0.047* |
| **Rand** | 0.093 | 0.923 | 21.064 | 0.663 | 1.417* | 62.686* | 0.133* |
| **Swiss Franc** | 0.025 | 0.697 | 5.009 | 0.823 | 0.151 | 0.972* | 0.050* |

Notes: All exchange rates are in terms of one unit of a US Dollar. The Euro replaces the Deutsche Mark for the POST1999 sample period. Skew and Kurt represent the skewness and kurtosis coefficients respectively. Normality is tested using the Kolmogorov-Smirnov statistic. The critical value for the skewness, kurtosis and normality tests is zero. The symbol * indicates significance at the five percent level. Mean, standard deviation, range and interquartile range measures are presented in percentage form.

Figure 3: Plot of Daily Conditional Volatility for Euro Returns

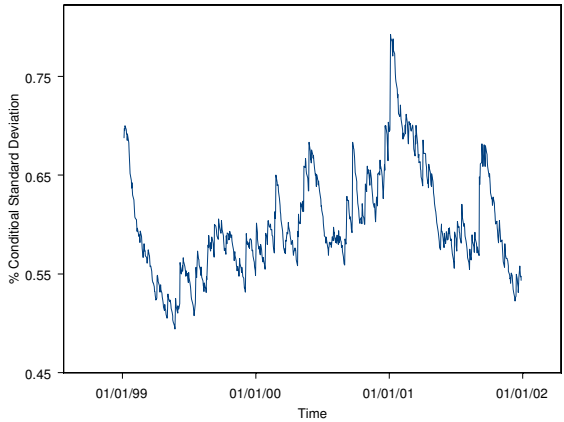

$$h^2_t \quad = \quad 0.006 \; + \quad 0.017\varepsilon^2_{t-i} \quad + \quad 0.96h^2_{t-j}$$
$$(0.856) \qquad (1.690)^* \qquad\qquad (34.381)^*$$

Notes: The figure details the daily percentage conditional volatility for the Euro based on a GARCH (1, 1) specification. The parameters and associated t-statistics in parenthesis of the GARCH specification are presented beneath the figure. The t-statistics use Bollerslev and Wooldridge (1992) standard errors. The symbol * denotes significant at the 5% level. The time-varying feature of the figure demonstrates volatility clustering with periods of high (low) volatility followed by further periods of high (low) volatility.

Table 3: Optimal Tail Estimates for Exchange Rate Returns Series

| Currency | $\gamma^{-}$ | $\gamma^{+}$ | $\gamma^{*}$ | $\gamma^{+}$ - $\gamma^{-}$ | $\gamma^{*}$ = 2 | $\gamma^{*}$ = 4 |
|---|---|---|---|---|---|---|
| **PRE1999** | | | | | | |
| **Deutsche Mark** | 3.51 | 3.40 | 3.82 | -0.24 | 2.63 | -0.26 |
| | (0.32) | (0.31) | (0.35) | | | |
| **Yen** | 2.49 | 3.47 | 3.05 | 2.48 | 1.90 | -1.72 |
| | (0.23) | (0.32) | (0.28) | | | |
| **Sterling** | 3.16 | 2.45 | 3.20 | -1.92 | 2.07 | -1.38 |
| | (0.29) | (0.23) | (0.30) | | | |
| **Rand** | 2.31 | 2.04 | 2.16 | -0.95 | 0.41 | -4.70 |
| | (0.21) | (0.19) | (0.20) | | | |
| **Swiss Franc** | 3.28 | 3.33 | 3.66 | 0.12 | 2.50 | -0.51 |
| | (0.30) | (0.31) | (0.34) | | | |
| | | | | | | |
| **POST1999** | | | | | | |
| **Euro** | 2.54 | 3.43 | 3.71 | 1.84 | 2.08 | -0.35 |
| | (0.29) | (0.39) | (0.42) | | | |
| **Yen** | 2.33 | 3.15 | 2.82 | 1.85 | 1.31 | -1.89 |
| | (0.26) | (0.36) | (0.32) | | | |
| **Sterling** | 2.72 | 3.30 | 3.20 | 1.20 | 1.69 | -1.13 |
| | (0.31) | (0.37) | (0.36) | | | |
| **Rand** | 2.40 | 1.78 | 2.36 | -1.83 | 0.69 | -3.13 |
| | (0.27) | (0.20) | (0.27) | | | |
| **Swiss Franc** | 2.40 | 3.20 | 3.40 | 1.77 | 1.86 | -0.80 |
| | (0.27) | (0.36) | (0.38) | | | |

Notes: The Euro replaces the Deutsche Mark for the POST1999 sample period. Hill tail estimates, $\gamma$, are calculated for lower, upper and both tails for each currency. The symbols -, +, * represent the lower, upper and both tails respectively. Standard errors are presented in parenthesis for each tail value. Tail stability is calculated in the third last column measuring divergence between upper and lower tail values with a critical value of 1.96. The second last and last columns determine whether the tail estimates, $\gamma^{*}$, are significantly different from 2 and 4 respectively with a critical value of 1.64.

Table 4: The Five Highest and Lowest Extreme Exchange Rate Returns

| Currency | | 1 | 2 | 3 | 4 | 5 |
|---|---|---|---|---|---|---|
| **PRE1999** | | | | | | |
| **Deutsche Mark** | *Highest* | 3.46 | 3.32 | 3.12 | 2.82 | 2.80 |
| | *Lowest* | -3.51 | -2.93 | -2.80 | -2.67 | -2.60 |
| **Yen** | *Highest* | 4.14 | 3.46 | 3.31 | 3.26 | 2.69 |
| | *Lowest* | -7.69 | -5.42 | -4.96 | -4.78 | -3.67 |
| **Sterling** | *Highest* | 4.29 | 3.47 | 3.42 | 2.72 | 2.59 |
| | *Lowest* | -4.22 | -3.42 | -3.10 | -3.07 | -2.91 |
| **Rand** | *Highest* | 3.59 | 3.48 | 3.40 | 3.35 | 3.35 |
| | *Lowest* | -4.44 | -4.12 | -3.14 | -2.62 | -2.54 |
| **Swiss Franc** | *Highest* | 3.78 | 3.63 | 3.58 | 2.97 | 2.66 |
| | *Lowest* | -3.71 | -3.30 | -3.26 | -3.10 | -3.02 |
| | | | | | | |
| **POST1999** | | | | | | |
| **Euro** | *Highest* | 2.55 | 2.30 | 1.89 | 1.86 | 1.76 |
| | *Lowest* | -2.36 | -2.26 | -2.07 | -2.03 | -2.02 |
| **Yen** | *Highest* | 3.95 | 3.38 | 2.22 | 2.05 | 1.88 |
| | *Lowest* | -2.85 | -2.84 | -2.79 | -2.56 | -2.36 |
| **Sterling** | *Highest* | 1.65 | 1.65 | 1.51 | 1.40 | 1.37 |
| | *Lowest* | -1.84 | -1.47 | -1.46 | -1.38 | -1.36 |
| **Rand** | *Highest* | 8.28 | 7.47 | 5.36 | 4.22 | 2.99 |
| | *Lowest* | -12.78 | -2.57 | -2.51 | -2.05 | -1.74 |
| **Swiss Franc** | *Highest* | 2.32 | 2.26 | 2.19 | 2.02 | 2.01 |
| | *Lowest* | -2.69 | -2.60 | -2.00 | -1.98 | -1.97 |

Notes: The table indicates the five highest and lowest percentage exchange rate first differences for each currency during the period of analysis. The Euro replaces the Deutsche Mark for the POST1999 sample period.

Table 5: Stability of Tail Estimates across Subperiods

| Currency | Deutsche Mark | Yen | Sterling | Rand | Swiss Franc |
|---|---|---|---|---|---|
| **Euro** | **-0.20** | 1.30 | 0.99 | 3.33 | 0.09 |
| **Yen** | -2.10 | **-0.54** | -0.87 | 1.75 | -1.81 |
| **Sterling** | -1.23 | 0.33 | **0.00** | 2.51 | -0.93 |
| **Rand** | -3.30 | -2.58 | -2.11 | **0.60** | -3.02 |
| **Swiss Franc** | -0.80 | 0.73 | 0.41 | 2.86 | **-0.51** |

Notes: The Euro replaces the Deutsche Mark for the POST1999 sample period. Hill tail estimates, $\gamma^*$, are compared across subperiods for PRE1999 and POST1999. Values represent measures of the stability test proposed by Loretan and Phillips (1994) calculated for divergence between PRE1999 to POST1999 with a critical value of 1.96. Positive (negative) values represent an increase (decrease) in tail estimates over time. The diagonal of the matrix represent the test statistic for divergence across a currency whereas non-diagonal values represent divergence across currencies.

Table 6: Tail Quantile Estimates for Exchange Rate Returns Series

| Currency | $r_p (95)$ | $r_p (99)$ | $r_p (99.5)$ | $r_p (1/n)$ | $r_p (1/2n)$ | $r_p (1/4n)$ |
|---|---|---|---|---|---|---|
| **PRE1999** | | | | | | |
| **Deutsche Mark** | 1.24 | 1.90 | 2.27 | 4.33 | 5.20 | 6.23 |
| **Yen** | 1.24 | 2.11 | 2.64 | 5.93 | 7.44 | 9.34 |
| **Sterling** | 1.07 | 1.77 | 2.20 | 4.76 | 5.91 | 7.33 |
| **Rand** | 0.94 | 1.97 | 2.72 | 8.51 | 11.73 | 16.17 |
| **Swiss Franc** | 1.33 | 2.07 | 2.50 | 4.90 | 5.92 | 7.15 |
| | | | | | | |
| **POST1999** | | | | | | |
| **Euro** | 1.15 | 1.78 | 2.14 | 3.09 | 3.72 | 4.49 |
| **Yen** | 1.06 | 1.88 | 2.40 | 3.88 | 4.97 | 6.35 |
| **Sterling** | 0.82 | 1.36 | 1.68 | 2.57 | 3.20 | 3.97 |
| **Rand** | 1.07 | 2.13 | 2.85 | 5.07 | 6.81 | 9.13 |
| **Swiss Franc** | 1.15 | 1.84 | 2.25 | 3.36 | 4.12 | 5.06 |

Notes: The Euro replaces the Deutsche Mark for the POST1999 sample period. $r_p(95)$, $r_p (99)$, and $r_p (99.5)$ represent the 95, 99, and 99.5 quantile confidence intervals respectively. These estimates measure the size of the currency fluctuations needed to cover the associated confidence intervals, for example, $r_p(95)$ represents 95% of all movements. $r_p (1/n)$, $r_p (1/2n)$, and $r_p (1/4n)$ deal with out-of-sample, n, exchange rate movements. These extrapolated estimates allow us to forecast the size of the currency fluctuations over longer periods of analysis than the data set used, for example, $r_p (1/4n)$ gives estimates for 9392 days (2348 * 4) for the PRE1999 sample period. All values are expressed in percentages.

Table 7: Tail Excess Probability Estimates for Exchange Rate Returns Series

| Currency | $P_r5\%$ | $P_r3\%$ | $P_r2.5\%$ | $P_r2\%$ | $P_r1.5\%$ | $P_r1\%$ |
|---|---|---|---|---|---|---|
| **PRE1999** | | | | | | |
| **Deutsche** | | | | | | |
| **Mark** | 0.02 | 0.17 | 0.35 | 0.82 | 2.45 | 11.55 |
| **Yen** | 0.07 | 0.34 | 0.59 | 1.17 | 2.82 | 9.70 |
| **Sterling** | 0.04 | 0.19 | 0.33 | 0.68 | 1.71 | 6.26 |
| **Rand** | 0.13 | 0.40 | 0.60 | 0.97 | 1.81 | 4.34 |
| **Swiss Franc** | 0.04 | 0.26 | 0.50 | 1.13 | 3.24 | 14.27 |
| | | | | | | |
| POST1999 | | | | | | |
| **Euro** | 0.02 | 0.14 | 0.28 | 0.64 | 1.87 | 8.43 |
| **Yen** | 0.06 | 0.27 | 0.44 | 0.83 | 1.88 | 5.90 |
| **Sterling** | 0.02 | 0.08 | 0.14 | 0.29 | 0.72 | 2.64 |
| **Rand** | 0.13 | 0.44 | 0.68 | 1.15 | 2.28 | 5.93 |
| **Swiss Franc** | 0.03 | 0.19 | 0.35 | 0.75 | 2.00 | 7.93 |

Notes: The Euro replaces the Deutsche Mark for the POST1999 sample period. $P_r5\%$, $P_{r3}\%$, $P_r2.5\%$, $P_r2\%$, $P_r1.5\%$, and $P_r1\%$ represent the probability of an exchange rate movement being in excess of the associated percentages, for example, 5 percent ($P_r5\%$). The excess probability estimates are presented in percentage form.

**<u>Appendix</u>**

Details of the limit laws and asymptotic convergence are outlined in this appendix. The limit laws of the maxima of n random variables belonging to a max-stable distribution given by the Fisher-Tippett theorem indicate that the maxima at the limit converges in distribution to H after normalising and centring. Formally this is expressed as

$$c_n^{-1}(M_n - d_n) \xrightarrow{d} H \qquad\qquad \text{for } c_n > 0, \ -\infty < d_n < \infty \qquad\qquad (A1)$$

Where $\xrightarrow{d}$ represents convergence in distribution, $c_n$ and $d_n$ are normalising and centring constants respectfully. In (A1) the asymptotic behaviour of $M_n$ is related (converges) to the asymptotic behaviour of the distribution H at its right tail as you are both dealing with the upper order statistics of the sequence.

The corresponding limit distribution for the maxima of a max-stable distribution is

$$c_n^{-1}(M_{n-}d_n) \underline{d} R \qquad\qquad \text{for } c_n > 0, \ -\infty < d_n < \infty \qquad\qquad (A2)$$

where $\underline{d}$ represents the sample maxima belonging to the same respective distribution. It can be said than any affine transformation of the max-stable distribution is the only limit distribution for the maxima of n random variables. The specific limit distributions for the three types of Extreme Value distributions are as follows

$$\text{Gumbell: } M_n \underline{d} R + \ln n$$
$$\text{Fréchet: } M_n \underline{d} n^{1/\alpha} R$$
$$\text{Weibull: } M_n \underline{d} n^{-1/\alpha} R \qquad\qquad (A3)$$

The tail shape parameter, $\alpha$, is related to the parameter $\gamma$. Using this notation, a single representation of the three Extreme Value distributions is outlined in the Generalised Extreme Value distribution and this is as follows:

$H_\gamma(r) \quad = \exp \left(-(1 + \gamma r)^{(-1/\gamma)}\right) \qquad$ if $\gamma \neq 0$, and

$\qquad = \exp \left(- \exp \left(-r\right)\right) \qquad$ if $\gamma = 0$, $\qquad$ (A4)

where $1 + \gamma r > 0$, and $\gamma = 0$ is to be regarded as the limit of the distribution function as $\gamma \to 0$. Equation (A4) is the Jenkinson-Von Mises representation of the generalised Extreme Value distribution.

This simplified representation of the three Extreme Value distributions focuses on the single parameter $\gamma$, which has the following relationship with $\alpha$:

Type I (Gumbell): $\Lambda$ $\qquad$ for $\gamma = 0$

Type II (Fréchet): $\Phi_\alpha$ $\qquad$ for $\gamma = \alpha^{-1} > 0$

Type III (Weibull): $\psi_\alpha$ $\qquad$ for $\gamma = -\alpha^{-1} < 0$

The necessary and sufficient conditions using Gnednenko's theorem for a distribution to asymptotically converge on a type of Extreme Value distributions are:

Type I (Gumbell): $\quad \lim_{t \to \infty} n \quad [1 - F(\alpha_n r + \beta_n)] \qquad = \qquad e^{-r}$

Type II (Fréchet): $\quad \lim_{t \to \infty} \dfrac{1 - F(tr)}{1 - F(t)} \quad = \quad r^{-1/\alpha} \quad = \quad r^{-\gamma}$

For $r > 0$, $\alpha > 0$.

Type III (Weibull): $\quad \lim_{t \to 0} \dfrac{1 - F(tr + u)}{1 - F(t + u)} \quad = \quad r^{1/\alpha} \quad = \quad r^{\gamma} \qquad$ (A5)

For $r > 0$, $\alpha < 0$.

**Footnotes:**

[1] More recently the Asian currency crises sent warning signals as to the stability of nominal exchange rates (for a review see Corsetti *et al*., 1998).

[2] These extrapolated estimates are described as out-of-sample estimates in the Extreme Value Theory literature.

[3] The Deutsche Mark is incorporated as a precursor to the Euro.

[4] Jensen's inequality says that for the exchange rate, s, expressed in foreign currency per unit of the local currency, $E(1/s) \neq 1/E(s)$. For example, whilst the DM/$ rate was the reciprical of the $/DM rate, their expected values were not reciprocals (Diebold, 1988).

[5] This may be due to investors perceiving that the Euro is a continuation of the German currency.

[6] The BHHH algorithm is used for convergence after the simplex method provides the initial starting values. The starting values for the conditional variance are given by the unconditional variance measures. The model was fitted assuming the conditional t-distribution to allow for fat-tails (Baillie and Bollerslev, 1990). AIC and BIC criteria were used for model selection.

[7] This interpretation is given in the context of comparing different exchange rate regimes, fixed and floating, using tail estimates (Koedijk *et al*., 1992)

[8] The Extreme Value estimates cover fully the final transitional period, 1999-2001, to a full monetary union.